# Gravitational interaction mediated by Classical Zero Point Field

Ion Simaciu[1,a], Zoltan Borsos[1,b], Viorel Drafta[2] and Gheorghe Dumitrescu[3]

[1] Retired lecturer, Petroleum-Gas University of Ploiești, Ploiești 100680, Romania

[2] Independent researcher

[3] High School Toma N. Socolescu, Ploiești, Romania

E-mail: [a] isimaciu@yahoo.com; [b] borzolh@upg-ploiesti.ro

**Abstract**

By modeling the particle as a two-dimensional oscillator with the natural angular frequency equal to the Zitterbewegung frequency, the expression of the gravitational force between two particles is obtained. Gravitational force is the effect of the absorption-scattering of the CZPF background by the oscillators. The connection between the gravitational and the electrostatic interaction is obtained.

1. **Introduction**

This paper proposes a model in which the gravitational interaction is mediated by CZPF radiation background. According to the model developed in the papers [1, 21] the electrostatic interaction between two charged particles, modeled as two-dimensional oscillators, is obtained as an effect of scattering the CZPF background.

Because the notion of scattering cross section [2, 21, 10] is used in the calculations, the intensity of the scattered radiation background is mediated in time and does not depend on the oscillation phases of the two oscillators. For this reason, the model does not allow the explanation that there are two types of electric charge.

To explain the property that both attractive and repulsive forces occur between oscillators it is necessary to take into account that the two oscillators can oscillate in phase (force is attractive) or in phase opposition (force is repulsive). The phenomenon is analogous to the interaction forces between two oscillating bubbles, i.e. the secondary Bjerknes forces [3, 4, 5, 17, 18].

Within the same model, the gravitational interaction between two particles can be explained if we accept the hypothesis that oscillators absorb some of the CZPF background radiation and transform it into a form of non-radiant energy (internal energy). This hypothesis is suggested by the theoretical and observational existence of stellar and galactic black holes [6]. The General Relativity highlights the existence of an electromagnetic radiation absorption cross section dependent on the angular frequency. The absorption cross section is proportional to the square of the gravitational radius (Schawrschild radius) for high frequencies [7, 8] and proportional to the fourth power of the gravitational radius and the square of the frequency for low frequencies [9].

At the microscopic level, the charged elementary particle is characterized by both a scattering cross section and an absorption cross section [10]. If the particle is at rest, the averaged electromagnetic scattering cross section [21] is proportional to the square of the electrostatic radius, i.e., the Thomson cross section [10, 1].

Although the particle is not collapsed at the gravitational radius (so it is not a mini black hole) it also has an absorption cross section dependent on angular frequency [10]. In the Appendix 2 we show that the averaged absorption cross section is proportional to the product between the electrostatic radius and the gravitational radius (Appendix 3).

Recent studies in Theoretical Physics have shown a link between the properties of elementary particles, the properties of black holes, and the properties of strings [11]. In the classical (non-quantum) context, we also record a connection between the discrete electric charge, relativity and the radiation thermodynamics [12].

In the charged particle model in Stochastic Electrodynamics [1, 21], the particle at rest is an oscillator that scatters the CZPF background and the scattering cross section depends on the angular frequency of the scattered wave and the natural angular frequency of the oscillator. If the oscillators are also characterized by the absorption of a fraction of the energy of the CZPF background, then between the two oscillators there is both a scattering force and a scattering-absorption force. We identify the scattering-scattering force as the electrostatic force [1, 21] we will prove that the scattering-absorption force is the gravitational force.

## 2. Gravitational force in EDS

We model a charged particle as a two-dimensional oscillator that scatters and absorbs CZPF background radiation. The oscillator with mass $m_i$ and the electric charge $q_i = |n_i q_e|$, ( $n_i = |\mp 1|, |\mp 1/3|, |\mp 2/3|, |q_e| = 1.602176634 \times 10^{-19}$ C ) is characterized by an interaction cross section with the plane electromagnetic wave [1, 10]

$$\sigma_{ti}(\omega) = \frac{4\pi R_i \omega^2 \Gamma_{ti} c}{\left[ \left( \omega_i^2 - \omega^2 \right)^2 + \Gamma_{ti}^2 \omega^2 \right]}. \tag{1}$$

In previous relationships, $R_i = q_i^2 / (4\pi\varepsilon_0 m_i c^2) = n_i^2 e^2 / (m_i c^2)$ is the electrostatic radius of the particle, $\Gamma_{ti}(\omega) = \Gamma_i'(\omega) + \omega^2 \Gamma_i / \omega_{0i}^2$ is the total decay constant, $\Gamma_i'(\omega)$ is the absorptive width, $\Gamma_i = \tau_i \omega_{0i}^2 = 2R_i \omega_{0i}^2 / (3c) = 2n_i^2 e^2 \omega_{0i}^2 / (3m_i c^3)$ is the radiative decay constant, $\omega_i$ is the natural angular frequency and $\omega$ the angular frequency of the plane electromagnetic wave [10].

According to Eq. (26) of paper [1, 21] the elementary force of interaction between two oscillators scattering the CZPF background has the expression

$$\delta F_{t12}(r,\omega) = \frac{\rho(\omega)\sigma_{t1}(\omega)\sigma_{t2}(\omega)d\omega}{3\pi r^2}. \tag{2}$$

The CZPF background is characterized by spectral density

$$\rho(\omega, T=0) = \frac{\hbar \omega^3}{2\pi^2 c^3}. \tag{3}$$

Replacing Eqs. (1) and (3) in Eq. (2) and integrating, results

$$F_{t12}(r) = \frac{8\hbar R_1 R_2}{3\pi r^2 c} \int_0^\Omega \frac{\Gamma_{t1}(\omega)\Gamma_{t2}(\omega)\omega^7 d\omega}{\left[ \left( \omega_1^2 - \omega^2 \right)^2 + \Gamma_{t2}^2(\omega)\omega^2 \right]\left[ \left( \omega_2^2 - \omega^2 \right)^2 + \Gamma_{t2}^2(\omega)\omega^2 \right]}, \Omega > \omega_2 \neq \omega_1 \tag{4}$$

Replacing the total decay constant $\Gamma_{ti}(\omega)$, Eq. (4) has the expression

$$F_{t12}(r) = \frac{8\hbar R_1 R_2}{3\pi r^2 c}\int_0^\Omega \frac{\left(\Gamma_1'(\omega)+\omega^2\Gamma_1/\omega_{01}^2\right)\left(\Gamma_2'(\omega)+\omega^2\Gamma_2/\omega_{02}^2\right)\omega^7 d\omega}{\left[\left(\omega_1^2-\omega^2\right)^2+\Gamma_{t1}^2(\omega)\omega^2\right]\left[\left(\omega_2^2-\omega^2\right)^2+\Gamma_{t2}^2(\omega)\omega^2\right]} =$$

$$\frac{8\hbar R_1 R_2}{3\pi r^2 c}\int_0^\Omega \frac{\left(\frac{\omega^4\Gamma_1\Gamma_2}{\omega_{01}^2\omega_{02}^2}+\frac{\omega^2\Gamma_1'(\omega)\Gamma_2}{\omega_{02}^2}+\frac{\omega^2\Gamma_1\Gamma_2'(\omega)}{\omega_{01}^2}+\Gamma_1'(\omega)\Gamma_2'(\omega)\right)\omega^7 d\omega}{\left[\left(\omega_1^2-\omega^2\right)^2+\Gamma_{t1}^2(\omega)\omega^2\right]\left[\left(\omega_2^2-\omega^2\right)^2+\Gamma_{t2}^2(\omega)\omega^2\right]} =$$

$$\frac{8\hbar R_1 R_2}{3\pi r^2 c}\int_0^\Omega \frac{\left[\Gamma_1\Gamma_2/\left(\omega_{01}^2\omega_{02}^2\right)\right]\omega^{11} d\omega}{\left[\left(\omega_1^2-\omega^2\right)^2+\Gamma_{t1}^2(\omega)\omega^2\right]\left[\left(\omega_2^2-\omega^2\right)^2+\Gamma_{t2}^2(\omega)\omega^2\right]} +$$

$$\frac{8\hbar R_1 R_2}{3\pi r^2 c}\int_0^\Omega \frac{\Gamma_1'(\omega)\Gamma_2/\omega_{02}^2\,\omega^9 d\omega}{\left[\left(\omega_1^2-\omega^2\right)^2+\Gamma_{t1}^2(\omega)\omega^2\right]\left[\left(\omega_2^2-\omega^2\right)^2+\Gamma_{t2}^2(\omega)\omega^2\right]} +$$

$$\frac{8\hbar R_1 R_2}{3\pi r^2 c}\int_0^\Omega \frac{\left[\Gamma_1\Gamma_2'(\omega)/\omega_{01}^2\right]\omega^9 d\omega}{\left[\left(\omega_1^2-\omega^2\right)^2+\Gamma_{t1}^2(\omega)\omega^2\right]\left[\left(\omega_2^2-\omega^2\right)^2+\Gamma_{t2}^2(\omega)\omega^2\right]} +$$

$$\frac{8\hbar R_1 R_2}{3\pi r^2 c}\int_0^\Omega \frac{\left(\Gamma_1'(\omega)\Gamma_2'(\omega)\right)\omega^7 d\omega}{\left[\left(\omega_1^2-\omega^2\right)^2+\Gamma_{t1}^2(\omega)\omega^2\right]\left[\left(\omega_2^2-\omega^2\right)^2+\Gamma_{t2}^2(\omega)\omega^2\right]} =$$

$$\frac{8\hbar R_1 R_2}{3\pi r^2 c}I_{s\Omega}+\frac{8\hbar R_1 R_2}{3\pi r^2 c}\left(I_{sa\Omega 1}+I_{sa\Omega 2}\right)+\frac{8\hbar R_1 R_2}{3\pi r^2 c}I_{a\Omega},\ \Omega>\omega_2\neq\omega_1. \tag{5}$$

In the final force relation in Eq. (5), the first term $8\hbar R_1 R_2 I_{s\Omega}/(3\pi r^2 c)$, represents the expression of the Coulomb force (scattering-scattering force). The second term and the third term $8\hbar R_1 R_2\left(I_{sa\Omega 1}+I_{sa\Omega 2}\right)/(3\pi r^2 c)$ represent the expression of the scattering-absorption force and the fourth term $8\hbar R_1 R_2 I_{a\Omega}/(3\pi r^2 c)$ the expression of the absorption-absorption force.

In our universe, the force of gravitational interaction between fundamental particles is mainly the force between protons, protons and neutrons and between neutrons. In order to be able to apply our model to both the interaction between protons and neutrons and between neutrons it is necessary to accept that the interaction takes place between the charged particles that are components of protons and neutrons. These components are quarks: the up and down quarks [13]. The quarks are characterized by two types of mass [13, 14]: the current/naked/bare quark mass ($m_{un}=1.8-2.8\,\text{MeV}/c^2 = 3.52-5.48\,m_e$, $n_1=n_u=2/3$; $m_{dn}=4.3-5.2\,\text{MeV}/c^2 = 8.41-10.17\,m_e$, $n_2=n_d=-1/3$) and the constituent quark mass ($m_{dc}=340\,\text{MeV}/c^2 = 665.36\,m_e$, $m_{uc}=336\,\text{MeV}/c^2 = 657.53\,m_e$). The three quarks, with almost equal constituent masses ($m_{qc}\cong 338\,\text{MeV}/c^2 \cong 660\,m_e$), interact through strong attractive forces (potential energy is negative in the order of tens of MeV/c$^2$) so that the mass of the particle formed (proton and neutron or nucleons) becomes $m_{p/n}\cong 940\,\text{MeV}/c^2 \cong 1840\,m_e$. For this reason, we will consider that the particles that interact by scattering-absorption have approximately equal masses and different electrical charges. This approximation allows an analytical estimation of the expression of the gravitational force as the scattering-absorption force of the CZPF background.

According to Eq. (5), the scattering-absorption force has the expression

$$F_{sa12}(r) = \frac{8\hbar R_1 R_2}{3\pi r^2 c}\left[\int_0^\Omega \frac{\Gamma'_1(\omega)\Gamma_2/\omega_{02}^2 \, \omega^9 d\omega}{\left[(\omega_1^2 - \omega^2)^2 + \Gamma_{t1}^2(\omega)\omega^2\right]\left[(\omega_2^2 - \omega^2)^2 + \Gamma_{t2}^2(\omega)\omega^2\right]} + \right.$$

$$\left. \int_0^\Omega \frac{\left[\Gamma_1\Gamma'_2(\omega)/\omega_{01}^2\right]\omega^9 d\omega}{\left[(\omega_1^2 - \omega^2)^2 + \Gamma_{t1}^2(\omega)\omega^2\right]\left[(\omega_2^2 - \omega^2)^2 + \Gamma_{t2}^2(\omega)\omega^2\right]}\right], \Omega > \omega_2 \neq \omega_1. \quad (6)$$

If we approximate the total decay constant $\Gamma_{ti}(\omega) = \Gamma'_i + (\omega^2/\omega_i^2)\Gamma_i \cong (\omega^2/\omega_i^2)\Gamma_i = 2R_i\omega^2/(3c)$ because the absorptive width (absorptive decay constant) is $\Gamma'_i \ll \Gamma_i(\omega^2/\omega_i^2)$, the expression of the scattering-absorption force (6), becomes

$$F_{sa12}(r) = \frac{8\hbar R_1 R_2}{3\pi r^2 c}\left[\int_0^\Omega \frac{\Gamma'_1(\omega)(2R_2/(3c))\omega^9 d\omega}{\left[(\omega_1^2 - \omega^2)^2 + (2R_1/(3c))^2 \omega^6\right]\left[(\omega_2^2 - \omega^2)^2 + (2R_2/(3c))^2 \omega^6\right]} + \right.$$

$$\left. \int_0^\Omega \frac{\Gamma'_2(\omega)(2R_1/(3c))\omega^9 d\omega}{\left[(\omega_1^2 - \omega^2)^2 + (2R_1/(3c))^2 \omega^6\right]\left[(\omega_2^2 - \omega^2)^2 + (2R_2/(3c))^2 \omega^6\right]}\right] = \quad (7)$$

$$\frac{8\hbar R_1 R_2}{3\pi r^2 c}(I_{sa\Omega 1} + I_{sa\Omega 2}), \Omega > \omega_2 \neq \omega_1.$$

In the case of almost equal masses (since $m_2 \cong m_1 = m_{12}$ and the natural frequencies are proportional to the masses, it turns out $\omega_2 \cong \omega_1 = \omega_{12}$), the expression of force (7) becomes

$$F_{sa12}(r) = \frac{8\hbar R_1 R_2}{3\pi r^2 c}\left[\int_0^\Omega \frac{\Gamma'_1(\omega)(2R_2/(3c))\omega^9 d\omega}{\left[(\omega_{12}^2 - \omega^2)^2 + (2R_1/(3c))^2 \omega^6\right]\left[(\omega_{12}^2 - \omega^2)^2 + (2R_2/(3c))^2 \omega^6\right]} + \right.$$

$$\left. \int_0^\Omega \frac{\Gamma'_2(\omega)(2R_1/(3c))\omega^9 d\omega}{\left[(\omega_{12}^2 - \omega^2)^2 + (2R_1/(3c))^2 \omega^6\right]\left[(\omega_{12}^2 - \omega^2)^2 + (2R_2/(3c))^2 \omega^6\right]}\right] = \quad (8)$$

$$\frac{8\hbar R_1 R_2}{3\pi r^2 c}(I_{sa\Omega 1} + I_{sa\Omega 2}), \Omega > \omega_2 \cong \omega_1 = \omega_{12}.$$

The integrals $I_{sa1\Omega}$ and $I_{sa2\Omega}$ are analytically approximated by the saddle point method in Appendix 1. Replacing $I_{sa1\Omega}$ and $I_{sa2\Omega}$ integrals given by Eqs. (A1.3) and (A1.7), in the expression of force (8), results

$$F_{sa12}(r) = \frac{8\hbar R_1 R_2}{3\pi r^2 c}(I_{\Omega 1} + I_{\Omega 2}) = \frac{8\hbar R_1 R_2}{3\pi r^2 c}\left(\frac{3^2 \pi c^2 \Gamma'_1(\omega_{12})}{2^3 \omega_{12} R_1 (R_1 + R_2)} + \frac{3^2 \pi c^2 \Gamma'_2(\omega_{12})}{2^3 \omega_{12} R_2 (R_1 + R_2)}\right) =$$
$$\frac{3\hbar c R_1 R_2}{\omega_{12} r^2 (R_1 + R_2)}\left(\frac{\Gamma'_1(\omega_{12})}{R_1} + \frac{\Gamma'_2(\omega_{12})}{R_2}\right) = \frac{3\hbar c}{r^2} \frac{R_2 \Gamma'_1(\omega_{12}) + R_1 \Gamma'_2(\omega_{12})}{\omega_{12}(R_1 + R_2)}. \qquad (9)$$

Replacing the expressions of the electrostatic radii $R_i = n_i^2 e^2/(m_i c^2), i = 1, 2$ in Eq. (9), result

$$F_{sa12}(r) = \frac{3\hbar c}{r^2} \frac{R_2 \Gamma'_1(\omega_{12}) + R_1 \Gamma'_2(\omega_{12})}{\omega_{12}(R_1 + R_2)} = \frac{3\hbar c}{r^2} \frac{n_2^2 m_1 \Gamma'_1(\omega_{12}) + n_1^2 m_2 \Gamma'_2(\omega_{12})}{\omega_{12}(n_1^2 m_2 + n_2^2 m_1)}. \qquad (10)$$

To obtain the mass-dependent formula of the two particles it is necessary to calculate the expression of the absorptive width $\Gamma'_i(\omega)$. To achieve this task we calculate the average absorption cross section $\langle \sigma_{sai} \rangle_\omega$. According to the calculations in Appendix 2, the expression of the average absorption cross section is

$$\langle \sigma_{sai} \rangle_\omega = \frac{12\pi^2 c^2 \omega_i \Gamma'_i(\omega_i)}{\Omega^4}, \ \Omega > \omega_i. \qquad (11)$$

Replacing in Eq. (11) the maximum angular frequency, $\Omega(\omega_i) = \omega_i (3\pi c/(\omega_i R_i))^{1/4}$ (Eq. (17) in the paper [21]), corresponding to the averaged scattering cross section, is obtained

$$\langle \sigma_{sai} \rangle_\omega = \frac{12\pi^2 c^2 \omega_i \Gamma'(\omega_i)}{\Omega^4} = \frac{4\pi c R_i \Gamma'(\omega_i)}{\omega_i^2}. \qquad (12)$$

This section is independent of the natural frequency $\omega_i$ if $\Gamma'(\omega_i) \sim \omega_i^2$, i.e. $\Gamma'(\omega_i) = \tau' \omega_i^2$ and so

$$\langle \sigma_{sa} \rangle_\omega = 4\pi c R_i \tau'. \qquad (13)$$

For the section to have the dimensions of a surface, the constant $\tau'$ it is necessary to have the expression $\tau' = a'(R'_i/c)$ with $a'$ numerical constant (dimensionless). Replacing the constant $\tau'$ in Eq. (13), is obtained

$$\langle \sigma_{sa} \rangle_\omega = 4\pi a' c R_i R'_i. \qquad (14)$$

With the expression $\tau' = a'(R'_i/c)$, $\Gamma'(\omega_i)$ gets

$$\Gamma'(\omega_i) = \tau' \omega_i^2 = a' \frac{R'_i \omega_i^2}{c}, \qquad (15)$$

which is analogous to the expression of the radiative decay $\Gamma_i = \tau_i \omega_i^2 = 2R_i \omega_i^2/(3c)$. This analogy suggests that the radius $R'_i$ is, similar to the electromagnetic radius $R_i$, gravitational radius.

According to the gravitoelectromagnetic theory of the gravitational field [15] this radius is the minimum radius for which the particle energy is equal to the modulus of the energy of the gravitational field (Appendix 3). The energy of the gravitational field is negative because it corresponds to the phenomenon of absorption of the CZPF background energy by the particle which leads to a decrease of the average background density from zero point field. The absorption determines that the Poynting vector for the gravitational field is oriented towards the particle [15]. In general relativity, bodies in accelerated motion emit energy in the form of gravitational radiation [16-Sch. 105]. In accordance with the results of the third section, the expression of the radiated power, corresponding to the two-dimensional microscopic oscillator (fundamental particle), has the same shape as the power absorbed from the CZPF background.

Replacing the expression (A3.4) of the radius $R_i'$ in Eq. (15), result

$$\Gamma_i'(\omega_i) = a' \frac{G m_i \omega_i^2}{c^3}. \tag{16}$$

Replacing the expressions in Eq. (16) in Eq. (10), we obtain

$$F_{sa12}(r) = \frac{3\hbar c}{r^2} \frac{a' G \omega_{12}}{c^3} \frac{n_2^2 m_1^2 + n_1^2 m_2^2}{(n_2^2 m_1 + n_1^2 m_2)} = \frac{G m_1 m_2}{r^2} \frac{3 a' \hbar \omega_{12} (n_2^2 m_1^2 + n_1^2 m_2^2)}{c^2 m_1 m_2 (n_2^2 m_1 + n_1^2 m_2)}. \tag{17}$$

Replacing the natural angular frequency, $\omega_{12} = \omega_0 = c^2 \left( n_1^2 m_2 + n_2^2 m_1 \right) / \left( 2\hbar n_1 n_2 \right)$ (Eq, (11) in the paper [21]) deduced for electrostatic force, we obtain

$$F_{sa12}(r) = \frac{G m_1 m_2}{r^2} \frac{3 a' \left( n_2^2 m_1^2 + n_1^2 m_2^2 \right)}{2 n_1 n_2 m_1 m_2} \cong \frac{G m_1 m_2}{r^2} \frac{3 a' \left( n_2^2 + n_1^2 \right)}{2 n_1 n_2}. \tag{18}$$

This expression of the absorption force is proportional to the expression of the gravitational force $F_{g12}(r) = G m_1 m_2 / r^2$. This interaction is attractive, regardless of the sign of the electric charge (independent of the phase difference of the oscillators: in phase or phase opposition), because it is the result of energy absorption by one oscillator of the energy scattered from the CZPF background by the other oscillator it interacts with.

This demonstrates the connection between electromagnetism, gravitation and the CZPF background. The result is the same for any particle with mass $m$ and charge $q$ as stated by Boyer [12].

### 3. The power absorbed from the CZPF background

The expression of the absorbed power from the CZPF background [1] is

$$P_{sai} = \frac{2c}{3} \int_0^\Omega \rho(\omega) \sigma_{ai}(\omega) d\omega, \Omega > \omega_i \tag{19}$$

Replacing the absorption cross section (11), results

$$\begin{aligned} P_{sai} &= \frac{2c}{3} \int_0^\Omega \frac{\hbar \omega^3}{2\pi^2 c^3} \frac{4\pi c R_i \omega^2 \Gamma' d\omega}{\left[ \left( \omega_i^2 - \omega^2 \right)^2 + \left( 2 R_i / (3c) \right)^2 \omega^6 \right]} = \\ &\quad \frac{4 \hbar R_i}{3\pi c} \int_0^\Omega \frac{4 \omega^5 \Gamma' d\omega}{\left[ \left( \omega_i^2 - \omega^2 \right)^2 + \left( 2 R_i / (3c) \right)^2 \omega^6 \right]} = \\ &\quad \frac{\left( -\hbar R_i \omega_i^3 \Gamma'(\omega_i) \right)}{3\pi c} \int_0^\Omega \frac{dx}{\left[ x^2 + \left( R_i \omega_i^2 / (3c) \right)^2 \right]} = \\ &\quad \frac{\left( -\hbar R_i \omega_i^3 \Gamma'(\omega_i) \right)}{3\pi c} \left( \frac{-3\pi c}{R_i \omega_i^2} \right) = \hbar \omega_i \Gamma'(\omega_i), \omega_i - \omega = x, \Omega > \omega_i. \end{aligned} \tag{20}$$

Replacing $\Gamma'(\omega_i) = \tau' \omega_i^2 = a' (R_i'/c) \omega_i^2$ in Eq. (20), result

$$P_{sai} = \hbar\omega_i \Gamma'(\omega_i) = a'\frac{\hbar R'_i \omega_i^3}{c}, \qquad (21)$$

which is analogous to the expression of scattered power $P_{si} = 2\hbar R_i \omega_i^3/(3c) \gg P_{sai}$ [1, 22]. The energies corresponding to these powers are involved in the interactions between the particles considered as the first energy extractor from ZPF background [22].

If we replace angular natural frequency $\omega_i \cong m_i c^2/\hbar$ and gravitational radius $R'_i = Gm_i/c^2$ in Eq. (21), result

$$P_{sai} = a'\frac{Gm_i^4 c^4}{\hbar^2}, \qquad (22)$$

The absorbed power increases the rest mass of the particle. Increasing the mass involves increasing the internal temperature of the particle [23, 24]. The phenomenon is analogous to the increase in the temperature of the gas in the bubble and the fluid around an oscillating bubble by absorbing the energy of the acoustic waves [4, 5]. The law of variation in time is obtained by solving the equation (the fundamental constants do not change over time)

$$\frac{c^2 dm_i}{dt} = a'\frac{Gm_i^4 c^4}{\hbar^2} \text{ or } \frac{dm_i}{m_i^4} = a'\frac{Gc^2}{\hbar^2}dt. \qquad (23)$$

By integrating Eq. (23) we obtain

$$\int_{m_{i0}}^{m_i} \frac{dm_i}{m_i^4} = a'\frac{Gc^2}{\hbar^2}\int_0^t dt \text{ or } \frac{1}{m_i^3} = \frac{1}{m_{i0}^3} - a'\frac{3Gc^2}{\hbar^2}t, \; m_{i0} = m_i(t=0). \qquad (24)$$

The absorbed power is very small. The time in which the absorbed energy doubles the rest energy of a particle, $m_{i0} = m_i/2$, is

$$T_2 = \frac{7\hbar^2}{3a'Gc^2 m_i^3}. \qquad (25)$$

For the electron $m_e \cong 10^{-30}$ kg, this time is longer than the age of the universe $T_2 \cong 10^{23}$ s $\gg T_U \cong 10^{18}$ s.

The dependence of the absorbed power (21) on the gravitational radius and the third power of the angular frequency is also confirmed by the expression of the power radiated by a microscopic oscillator calculated according to general relativity. To obtain the expression of the radiated power for the two-dimensional microscopic oscillator, we replace in the expression of the gravitational power radiated by a system with the reduced mass $m_i$ which rotates with the angular velocity $\omega_i$ on a radius orbit $r_i$ [16 - Sch. 105]

$$P_{ri} = \frac{32Gm_i^2 r_i^4 \omega_i^6}{5c^5}. \qquad (26)$$

This system is equivalent to a two-dimensional microscopic oscillator [1, 21] for which $r_i\omega_i = c$ and $r_i = \hbar/(m_i c)$. With these relations, the expression of the radiated power (26) becomes

$$P_{ri} = \frac{32Gm_i^2 (r_i\omega_i)^3 r_i \omega_i^3}{5c^5} = \frac{32Gm_i^2 r_i \omega_i^3}{5c^2} = \frac{32}{5}\left(\frac{Gm_i}{c^2}\right)\frac{\hbar\omega_i^3}{c}. \qquad (27)$$

The fundamental difference is that this power is radiated in the form of gravitational waves [16]. For the energy balance of the microscopic system (microscopic particle), it must absorb energy from the outside with a power equal to the radiated power. Although it is accepted that the particle has an internal oscillating motion (Zitterbewegung) [19] and implicitly an internal acceleration its implications on the particle model have not been studied..

## 4. Conclusions

This paper highlights the connection between the electrostatic interaction and the gravitational interaction within SED. The electrostatic interaction between two oscillators is the effect of the reciprocal scattering of the secondary radiation generated by the primary scattering of the CZPF background radiation by them. The electric charge is a measure of the capacity of fundamental particles to scatter the CZPF radiation background.

This phenomenon has an analogy in the physics of the interaction between two oscillating bubbles in liquid [3, 4, 5, 17, 18]. The electrical acoustic charge is of two types because oscillators can oscillate in phase or phase opposition.

The gravitational interaction between the two oscillators is the effect of absorbing the secondary radiation generated by the primary scattering of the CZPF background radiation by them. This interaction is attractive because it is the effect of the absorption of the energy of the CZPF background by the two particles and its transformation into non-radiant energy (internal energy). Electrically neutral particles, but composed of charged particles with opposite charges as a sign, will only interact gravitationally. The interaction by absorbing the scattered radiation from the CZPF background is attractive, regardless of the sign of the charge of the component particles. The gravitational charge is a measure of the property of fundamental particles to absorb the scattered radiation from CZPF background.

The proposed model also has shortcomings. The Zitterbewegung condition obtained in classical (non-quantic) relativistic model of the electron [19] is $\hbar\omega_0 = 2mc^2$ and not the one obtained in the paper [1] and used to deduce the expression of gravitational force. Also, the gravitational radius $R_i'$ (A 3.4) is half the Schwarzschild radius [16 - Sch. 97]. These approximations determine the expression of electrostatic and gravitational forces up to a numerical constant. These shortcomings could be remedied in a rigorous approach in the SED for the interaction between two charged particles modeled as oscillators.

**Appendix 1. Absorption-scattering force**

With $\omega_2 \cong \omega_1 = \omega_{12}$, the expression of force (8) becomes

$$F_{sa12}(r) = \frac{8\hbar R_1 R_2}{3\pi r^2 c} \left[ \int_0^\Omega \frac{\Gamma_1'(\omega)(2R_2/(3c))\omega^9 d\omega}{\left[(\omega_{12}^2 - \omega^2)^2 + (2R_1/(3c))^2 \omega^6\right]\left[(\omega_{12}^2 - \omega^2)^2 + (2R_2/(3c))^2 \omega^6\right]} + \right.$$

$$\left. \int_0^\Omega \frac{\Gamma_2'(\omega)(2R_1/(3c))\omega^9 d\omega}{\left[(\omega_{12}^2 - \omega^2)^2 + (2R_1/(3c))^2 \omega^6\right]\left[(\omega_{12}^2 - \omega^2)^2 + (2R_2/(3c))^2 \omega^6\right]} \right] = \qquad (A1.1)$$

$$\frac{8\hbar R_1 R_2}{3\pi r^2 c}(I_{sa1\Omega} + I_{sa2\Omega}), \Omega > \omega_2 \cong \omega_1 = \omega_{12}.$$

The expression of the integral $I_{sa1\Omega}$ is

$$I_{sa1\Omega} = \int_0^\Omega \frac{\Gamma_1'(\omega)(2R_2/(3c))\omega^9 d\omega}{\left[(\omega_{12}^2-\omega^2)^2+(2R_1/(3c))^2\omega^6\right]\left[(\omega_{12}^2-\omega^2)^2+(2R_2/(3c))^2\omega^6\right]} =$$

$$\int_0^\Omega \frac{\Gamma_1'(\omega)(2R_2/(3c))\omega^9 d\omega}{\left[(\omega_{12}-\omega)^2(\omega_{12}+\omega)^2+\left(\frac{2R_1}{3c}\right)^2\omega^6\right]\left[(\omega_{12}-\omega)^2(\omega_{12}+\omega)^2+\left(\frac{2R_2}{3c}\right)^2\omega^6\right]} \cong$$

$$\int_{\omega_{12}}^{\omega_{12}-\Omega} \frac{\Gamma_1'(\omega_{12})(2R_2/(3c))\omega_{12}^9(-dx)}{\left[(\omega_{12}-\omega)^2(\omega_{12}+\omega)^2+\left(\frac{2R_1}{3c}\right)^2\omega_{12}^6\right]\left[(\omega_{12}-\omega)^2(\omega_{12}+\omega)^2+\left(\frac{2R_2}{3c}\right)^2\omega_{12}^6\right]} =$$

$$\frac{(-R_2\omega_{12}^5\Gamma_1'(\omega_{12}))}{2^3\cdot 3c}\int_{\omega_{12}}^{\omega_{12}-\Omega}\frac{dx}{\left[x^2+(R_1\omega_{12}^2/(3c))^2\right]\left[x^2+(R_2\omega_{12}^2/(3c))^2\right]} =$$

$$\frac{(-R_2\omega_{12}^5\Gamma_1'(\omega_{12}))}{2^3\cdot 3c}\int_{\omega_{12}}^{\omega_{12}-\Omega}\frac{dx}{x^4+x^2(\omega_{12}^2/(3c))^2(R_1^2+R_2^2)+\left[R_1R_2(\omega_{12}^2/(3c))^2\right]^2} =$$

$$\frac{(-R_2\omega_{12}^5\Gamma_1'(\omega_{12}))}{2^3\cdot 3c}I_{x1}, \quad \omega_{12}-\omega=x, \Omega>\omega_{12}. \tag{A1.2}$$

The $I_{sa1\Omega}$ integral is solved using formulas 2. 161 and 2.103 of the book [20]

$$I_{x1}=\int_{\omega_{12}}^{\omega_{12}-\Omega}\frac{dx}{x^4+x^2(\omega_{12}^2/(3c))^2(R_1^2+R_2^2)+\left[R_1R_2(\omega_{12}^2/(3c))^2\right]^2} =$$

$$\frac{2}{(\omega_{12}^2/(3c))^2(R_1^2-R_2^2)}\left[\int_{\omega_{12}}^{\omega_{12}-\Omega}\frac{dx}{2x^2+2R_2^2\left(\frac{\omega_{12}^2}{3c}\right)^2}-\int_{\omega_{12}}^{\omega_{12}-\Omega}\frac{dx}{2x^2+2R_1^2\left(\frac{\omega_{12}^2}{3c}\right)^2}\right] =$$

$$\frac{1}{(\omega_{12}^2/(3c))^2(R_1^2-R_2^2)}\left[\int_{\omega_{12}}^{\omega_{12}-\Omega}\frac{dx}{x^2+R_2^2\left(\frac{\omega_{12}^2}{3c}\right)^2}-\int_{\omega_{12}}^{\omega_{12}-\Omega}\frac{dx}{x^2+R_1^2\left(\frac{\omega_{12}^2}{3c}\right)^2}\right] = \tag{A1.3}$$

$$\frac{1}{(\omega_{12}^2/(3c))^2(R_1^2-R_2^2)}\left[\frac{3c}{R_2\omega_{12}^2}\arctan\frac{3cx}{R_2\omega_{12}^2}\bigg|_{\omega_{12}}^{\omega_{12}-\Omega}-\frac{3c}{R_1\omega_{12}^2}\arctan\frac{3cx}{R_1\omega_{12}^2}\bigg|_{\omega_{12}}^{\omega_{12}-\Omega}\right]\cong$$

$$\frac{(-3^3\pi c^3)}{\omega_{12}^6(R_1^2-R_2^2)}\left(\frac{R_1-R_2}{R_1R_2}\right)=\frac{(-3^3\pi c^3)}{\omega_{12}^6 R_1R_2(R_1+R_2)}, \frac{3c}{R_i\omega_{12}}\gg 1, \frac{3c\Omega}{R_i\omega_{12}^2}\gg 1, \Omega>\omega_{12}, i=1,2.$$

$$h = \left(\omega_{12}^2/(3c)\right)^4 \left(R_1^2 + R_2^2\right)^2 - 4\left[R_1 R_2 \left(\omega_{12}^2/(3c)\right)^2\right]^2 = \left(\omega_{12}^2/(3c)\right)^4 \left(R_1^2 - R_2^2\right)^2 > 0,$$

$$\left(\omega_{12}^2/(3c)\right)^2 \left(R_1^2 + R_2^2\right) - \sqrt{h} = 2R_2^2 \left(\omega_{12}^2/(3c)\right)^2, \quad \text{(A1.4)}$$

$$\left(\omega_{12}^2/(3c)\right)^2 \left(R_1^2 + R_2^2\right) + \sqrt{h} = 2R_1^2 \left(\omega_{12}^2/(3c)\right)^2, \text{ if } R_1 \neq R_2.$$

From the conditions $3c/(R_i \omega_{12}) \gg 1$ and $3c\Omega/(R_i \omega_{12}^2) \gg 1$, yields

$$\frac{3c}{R_i \omega_{12}} = \frac{3}{n_i^2} \frac{\hbar c}{e^2} \frac{m_i c^2}{\hbar \omega_{12}} \gg 1 \text{ sau } \frac{m_i c^2}{\hbar \omega_{12}} \gg \frac{n_i^2}{3} \frac{e^2}{\hbar c},$$

$$\frac{3c\Omega}{R_i \omega_{12}^2} = \frac{3}{n_i^2} \frac{\hbar c}{e^2} \frac{m_i c^2}{\hbar \omega_{12}} \frac{\Omega}{\omega_{12}} \gg 1 \text{ sau } \frac{m_i c^2}{\hbar \omega_{12}} \frac{\Omega}{\omega_{12}} \gg \frac{n_i^2}{3} \frac{e^2}{\hbar c}. \quad \text{(A1.5)}$$

Replacing Eq. (A1.3) in the expression of the integral (A1.2) we obtain

$$I_{sa1\Omega} = \frac{\left(-R_2 \omega_{12}^5 \Gamma_1'(\omega_{12})\right)}{2^3 \cdot 3c} I_{x1} = \frac{\left(-R_2 \omega_{12}^5 \Gamma_1'(\omega_{12})\right)}{2^3 \cdot 3c} \frac{\left(-3^3 \pi c^3\right)}{\omega_{12}^6 R_1 R_2 \left(R_1 + R_2\right)} =$$

$$\frac{3^2 \pi c^2 \Gamma_1'(\omega_{12})}{2^3 \omega_{12} R_1 \left(R_1 + R_2\right)}. \quad \text{(A1.6)}$$

The expression of the integral $I_{sa2\Omega}$ is

$$I_{sa2\Omega} = \int_0^\Omega \frac{\Gamma_2'(\omega) \left(2R_1/(3c)\right) \omega^9 d\omega}{\left[\left(\omega_{12}^2 - \omega^2\right)^2 + \left(2R_1/(3c)\right)^2 \omega^6\right] \left[\left(\omega_{12}^2 - \omega^2\right)^2 + \left(2R_2/(3c)\right)^2 \omega^6\right]} =$$

$$\int_0^\Omega \frac{\Gamma_2'(\omega) \left(2R_1/(3c)\right) \omega^9 d\omega}{\left[\left(\omega_{12} - \omega\right)^2 \left(\omega_{12} + \omega\right)^2 + \left(\frac{2R_1}{3c}\right)^2 \omega^6\right] \left[\left(\omega_{12} - \omega\right)^2 \left(\omega_{12} + \omega\right)^2 + \left(\frac{2R_2}{3c}\right)^2 \omega^6\right]} \cong$$

$$\int_{\omega_{12}}^{\omega_{12}-\Omega} \frac{\Gamma_2'(\omega) \left(2R_1/(3c)\right) \omega_{12}^9 (-dx)}{\left[\left(\omega_{12} - \omega\right)^2 \left(\omega_{12} + \omega\right)^2 + \left(\frac{2R_1}{3c}\right)^2 \omega_{12}^6\right] \left[\left(\omega_{12} - \omega\right)^2 \left(\omega_{12} + \omega\right)^2 + \left(\frac{2R_2}{3c}\right)^2 \omega_{12}^6\right]} =$$

$$\frac{\left(-R_2 \omega_{12}^5 \Gamma_1'(\omega_{12})\right)}{2^3 \cdot 3c} \int_{\omega_{12}}^{\omega_{12}-\Omega} \frac{dx}{\left[x^2 + \left(R_1 \omega_{12}^2/(3c)\right)^2\right] \left[x^2 + \left(R_2 \omega_{12}^2/(3c)\right)^2\right]} =$$

$$\frac{\left(-R_1 \omega_{12}^5 \Gamma_2'(\omega_{12})\right)}{2^3 \cdot 3c} \int_{\omega_{12}}^{\omega_{12}-\Omega} \frac{dx}{x^4 + x^2 \left(\omega_{12}^2/(3c)\right)^2 \left(R_1^2 + R_2^2\right) + \left[R_1 R_2 \left(\omega_{12}^2/(3c)\right)^2\right]^2} =$$

$$\frac{\left(-R_1 \omega_{12}^5 \Gamma_2'(\omega_{12})\right)}{2^3 \cdot 3c} I_{x2}, \, \omega_{12} - \omega = x, \Omega > \omega_{12}. \quad \text{(A1.7)}$$

The integrals $I_{x1}$ and $I_{x2}$ are identical and therefore

$$I_{sa1\Omega} = \frac{\left(-R_1\omega_{12}^5\Gamma'_2(\omega_{12})\right)}{2^3\cdot 3c}I_{x2} = \frac{\left(-R_1\omega_{12}^5\Gamma'_2(\omega_{12})\right)}{2^3\cdot 3c}I_{x1} =$$
$$\frac{\left(-R_1\omega_{12}^5\Gamma'_2(\omega_{12})\right)}{2^3\cdot 3c}\frac{\left(-3^3\pi c^3\right)}{\omega_{12}^6 R_1 R_2\left(R_1+R_2\right)} = \quad (A1.8)$$
$$\frac{3^2\pi c^2 \Gamma'_2(\omega_{12})}{2^3\omega_{12}R_2\left(R_1+R_2\right)}.$$

## Appendix 2. Averaged absorption cross section

The expression of averaged absorption cross section is

$$\left\langle\sigma_{sa}\right\rangle_\omega = \frac{\int_0^\Omega \sigma_{asi}(\omega)\rho(\omega)d\omega}{\int_0^\Omega \rho(\omega)d\omega}. \quad (A2.1)$$

Replacing in Eq. (A2.1) the expression of the absorption cross section $\sigma_{sai} \cong 4\pi c R_i \Gamma'_i \omega^2 / \left[\left(\omega_i^2 - \omega^2\right)^2 + \left(2R_i/(3c)\right)^2 \omega^6\right]$ [10], we obtain

$$\left\langle\sigma_{sa}\right\rangle_\omega = \frac{\int_0^\Omega\left[\frac{4\pi c R_i \Gamma'_i \omega^2}{\left[\left(\omega_i^2 - \omega^2\right)^2 + \left(2R_i/(3c)\right)^2 \omega^6\right]}\frac{\hbar\omega^3}{2\pi^2 c^3}\right]d\omega}{\int_0^\Omega \frac{\hbar\omega^3}{2\pi^2 c^3}d\omega} = \quad (A2.2)$$
$$\frac{4\int_0^\Omega \frac{4\pi c R_i \Gamma'_i \omega^5 d\omega}{\left[\left(\omega_i^2 - \omega^2\right)^2 + \left(2R_i/(3c)\right)^2 \omega^6\right]}}{\Omega^4}. \Omega > \omega_i.$$

To calculate the integral from the denominator, we apply the same method as for the integral (A1.3). In the integral from the denominator we make the change of variable $\omega_i - \omega = x$ and it becomes

$$I_{sa\Omega} = \int_0^\Omega \frac{4\pi c R_i \Gamma'_i \omega^5 d\omega}{\left[\left(\omega_i^2 - \omega^2\right)^2 + \left(2R_i/(3c)\right)^2 \omega^6\right]} = \int_0^\Omega \frac{4\pi c R_i \Gamma'_i \omega^5 d\omega}{\left[\left(\omega_i - \omega\right)^2\left(\omega_i + \omega\right)^2 + \left(2R_i/(3c)\right)^2 \omega_i^6\right]} \cong$$
$$4\pi c R_i \omega_i^5 \Gamma'_i(\omega_i)\int_{\omega_i}^{\omega_i-\Omega}\frac{(-dx)}{4\omega_i^2\left[x^2 + \left(\omega_i^2 R_i/(3c)\right)^2\right]} \cong \quad (A2.3)$$
$$\left(-\pi c R_i \omega_i^3 \Gamma'_i(\omega_i)\right)\left(\frac{-3\pi c}{\omega_i^2 R_i}\right) = 3\pi^2 c^2 \omega_i \Gamma'_i(\omega_i).$$

The integral, in the variable $x$, are solved according to the formulas 2.103 of the book [20, p. 67]. Replacing the integral (A2.3) in Eq. (A2.2), is obtained

$$\langle\sigma_{sa}\rangle_{\omega} = \frac{12\pi^2 c^2 \omega_i \Gamma'(\omega_i)}{\Omega^4}, \Omega > \omega_i. \quad (A2.4)$$

**Appendix 3. Gravitational radius**

The gravitational energy density, in Gravitoelectromagnetic theory [15], is

$$|w_{gi}| = \left|\frac{-1}{8\pi G}\left(E_g^2 + B_g^2\right)\right| = \frac{E_g^2}{4\pi G}, B_g^2 = E_g^2. \quad (A3.1)$$

With $E_{gi}^2 = G^2 m_i^2 / r^4$, result

$$|w_{gi}| = \frac{E_g^2}{4\pi} = \frac{Gm_i^2}{4\pi r^4}. \quad (A3.2)$$

The total energy of the field around the particle (modulus of this energy) is

$$|E_{gi}| = \int_{R_i'}^{\infty} |w_{gi}| 4\pi r^2 dr = Gm_i^2 \int_{R_i'}^{\infty} \frac{dr}{r^2} = \frac{Gm_i^2}{R_i'}. \quad (A3.3)$$

If we consider this as the rest energy of the particle $E_i = m_i c^2$, the expression of the gravitational radius is

$$m_i c^2 = \frac{Gm_i^2}{R_i'} \text{ sau } R_i' = \frac{Gm_i}{c^2}. \quad (A3.4)$$

This radius is half of the Schwarzschild radius [16 - Sch. 97].